# Distributed Agent System: Fault-Tolerant Collaboration Among Embodied Agents

Kai Yu*, Lu Chen, Hanqi Li
[a] X-LANCE Lab, School of Computer Science, Shanghai Jiao Tong University, Shanghai, 200240, China

[b] Jiangsu Key Lab of Language Computing, Suzhou, 215028, China

[*] Corresponding author. E-mail: kai.yu@sjtu.edu.cn, Tel: 13917575615, Address: 800 Dongchuan Road, Minhang District, Shanghai, 200240, China

**ABSTRACT**

AI engineering is shifting from passive text generation by large language models (LLMs) to agent-driven task execution, creating new reliability challenges for long-horizon tasks under resource constraints and environmental uncertainty. Conventional error-elimination optimization strategies fail to address cumulative error propagation. This paper proposes Distributed Agent System (DAS), a device-edge-cloud framework for fault-tolerant collaboration among heterogeneous agents. We redefine agent reliability as system-level fault tolerance rather than single-turn zero-error accuracy, and present a two-layer fault-tolerance architecture: single-agent execution reliability via fault-tolerant alignment, and cross-agent communication reliability via semi-formal language protocols. This framework provides a practical engineering pathway for reliable heterogeneous embodied agents collaboration in industrial scenarios.

## 1. Introduction: From Generative AI to Agentic Task Execution

AI engineering is undergoing a fundamental shift from passive text generation dominated by large language models (LLMs) towards agent-driven, end-to-end task execution [1]. Unlike earlier application modes that rely on staged human intervention or straightforward instruction matching, next-generation agents can complete a closed loop of environmental perception, reasoning, planning, decision-making, and physical execution, thereby supporting long-horizon autonomous operations in industrial settings [2,3]. This paradigm shift also changes how AI reliability should be evaluated. Q&A-based LLM interaction is typically evaluated by single-turn accuracy, whereas autonomous industrial embodied agents must be evaluated by system-level stability and reliability across continual multi-step tasks.

Long-horizon autonomous execution poses new reliability challenges and has become a core bottleneck for industrial AI deployment. End-deployed lightweight embodied agents face strict constraints on computing power, storage, and bandwidth, with limited model capacity compared to cloud frontier LLMs. Dynamic disturbances in open industrial environments lead to inherent uncertainty, which cannot be fully eliminated by model scaling or data iteration. Critically, minor local errors accumulate and amplify along long task chains, potentially causing overall task failure.

Current mainstream optimization methods, including tool-parameter calibration, model confidence estimation, and multi-turn self-reflection, can mitigate model hallucinations and execution errors only in single-turn or local scenarios [4,5,6,7]. They cannot suppress the systematic amplification of errors in long-horizon tasks. Meanwhile, most multi-agent collaboration research is built on homogeneous cloud-hosted LLMs for virtual scenarios, prioritizing single-turn accuracy, and cannot adapt to heterogeneous industrial end-device deployment and physical execution [8]. To bridge this gap, we propose a distributed fault-tolerant collaboration framework suitable for end-device embodied agents, replacing error elimination with system-level fault-tolerance management.

## 2. Distributed Agent System: Definition and Scope

**A Distributed Agent System (DAS) is a distributed framework for collective intelligence, built for heterogeneous end-edge-cloud agent deployment. It balances local autonomous operations on end devices with global coordination and scheduling on the cloud, and is tailored to collaborative tasks among heterogeneous agents characterized by long-horizon decision sequences, high reliability requirements and cost sensitivity constraints.** The primary deployment units of DAS are embodied

agents deployed on end devices, including industrial robots, environmental sensing and control equipment, IoT terminals, and other physical hardware that tightly couple intelligent decision-making with physical execution.

DAS employs layered coordination with distinct responsibilities. Lightweight end-device agents handle real-time perception and basic autonomous actions. The edge aggregates regional data and coordinates local tasks. The cloud supports global planning and optimization. Crucially, all tiers run in distributed mode and make autonomous decisions independently—not centrally dispatched by the cloud—achieving a dynamic balance suitable for complex industrial workflows.

DAS differs fundamentally from cloud LLM-based multi-agent systems across five core dimensions. In architecture, it is built on heterogeneous device-edge-cloud agents rather than homogeneous clusters of cloud foundation models. In terms of resource assumptions, it operates under strict end-device constraints on computation, storage, and communication, in sharp contrast to the abundant computing power and comprehensive knowledge bases available in cloud-native systems. Regarding optimization objectives, it prioritizes system-level long-horizon decision sequence reliability through proactive error accumulation management instead of pursuing perfect single-turn output accuracy. For core challenges, it tackles error accumulation and coupled risks between decision-making and physical execution, rather than the relatively low-risk tasks of question answering and software decomposition faced by cloud-based systems. Finally, in application scenarios, it focuses on physical long-horizon operations in industrial settings as opposed to pure virtual software environments [9,10].

## 3. Redefining Reliability: From Error Elimination to Fault-Tolerance Management

Insufficient reliability is the primary bottleneck preventing large-scale industrial deployment of embodied agents. Current academic approaches to LLM uncertainty generally follow the same logic: use larger models, more training data, high-precision confidence estimation, and iterative reflection to eliminate hallucinations and execution errors entirely. This logic may be feasible in compute-rich, cloud-hosted scenarios with short interaction cycles, such as question answering and software tool use. However, it does not fit resource-constrained industrial deployment on end devices with long-horizon autonomous execution, where the operating conditions are fundamentally different.

We start from a basic insight: **uncertainty introduced by models and environments is not a system defect that must be completely eradicated. Instead, it is an inherent property of human-AI and AI-AI interaction and physical operation under incomplete information; it need not and cannot be fully eliminated.** From the perspective of cognitive linguistics, ambiguity in natural language and in exchanged information is an important feature of efficient collaboration [11,12,13]. Humans resolve such ambiguity dynamically through multi-turn interaction and adapt to complex scenarios. This metacognitive regulatory mechanism has been used in POMDP-based spoken dialogue systems [21,22] and can be transferred to the optimization of agent systems.

Industrial end-device constraints further support this: hardware resource limits prevent unbounded model scaling; open environments involve unseen operating conditions; and long-horizon tasks amplify small errors into task failure. Therefore, reliability optimization for industrial agents must shift its core logic from "error elimination" to "uncertainty management and error-propagation suppression". Based on this insight, we redefine agent reliability as follows:

**Reliability is the comprehensive system-level capability of an agent to complete a closed-loop task process stably, consistently, and controllably under given task and environmental constraints, ensuring safe behavior, trustworthy outcomes, predictable decisions, and fault-tolerant handling of exceptions. It is essentially a utility trade-off between task-execution benefits and the costs of hardware, time, and human intervention.**

This definition differs from traditional evaluation frameworks: it no longer judges reliability by accuracy or success in a single interaction, but instead treats end-to-end fault-tolerance management, error-amplification suppression, and stable global task completion as the core criteria. Small local errors can be gradually resolved through multi-turn interaction, dynamic information updating, and the complementary capabilities of multiple nodes.

On the basis of this new reliability definition, DAS for industrial long-horizon operation should adopt a **two-layer fault-tolerance architecture** to manage local errors and suppress their propagation at the system level. The first layer is **single-agent execution fault tolerance**: fault-tolerant actions such as proactive refusal, task routing, and user clarification are used to manage execution errors in individual embodied agents and reduce harmful hallucinated outputs at the source. The second layer is **multi-agent communication fault tolerance**: standardized semi-formal protocols are used to unify cross-agent communication rules, block the cascading propagation of semantic deviations and information ambiguity, and introduce dynamic formal verification and auditing during execution. The two layers work together to systematically address reliability degradation in DAS.

## 4. Paradigm One: Reliable Single-Agent Execution via Fault-Tolerant Reliability Alignment

To address agents' inherent model uncertainty, tool-calling hallucinations, and incorrect task assessment, we propose a fault-tolerant reliability alignment paradigm. This paradigm reconstructs the optimization objectives to meet the fault-tolerance requirements of lightweight agents. Its core innovation is to acknowledge the inherent nature of model uncertainty, move away from the objective of absolute zero error, and achieve reliable execution under uncertain conditions through diverse fault-tolerant actions [4,5].

Traditional LLM often treats non-response in unknown scenarios and erroneous outputs as equivalent negative samples, thereby neglecting the value of adaptive fault-tolerant behavior under uncertainty. Consequently, the output action space of traditional LLM agents is usually dominated by concrete task-execution actions such as direct responses; few policies explicitly distinguish fault-tolerant communicative actions such as refusal, clarification, confirmation and task delegation etc. When such agents are deployed on lightweight end devices, they may easily produce outputs beyond their capability boundaries, inducing tool-calling hallucinations and capability overreach, and therefore leading to unreliable industrial operation.

The fault-tolerant reliability alignment paradigm reframes model optimization around a comprehensive utility trade-off: maximizing task usefulness while minimizing interaction cost. It accounts for end-device computing overhead, the cost of erroneous execution, and the cost of human intervention, making it better suited to the resource constraints of lightweight agents. Its central breakthrough is to expand the action space for uncertainty handling and break the binary logic of "answer or error." In addition to standard task-execution actions, the paradigm introduces an indecisive action space consisting of adaptive fault-tolerant actions such as proactive refusal, intent clarification, leading questions and cross-layer task routing. It also adopts an end-to-end trainable optimization framework and uses reinforcement learning for continuous iteration. This distinguishes it from passive "confidence detection plus fixed rules" methods and enables data-driven dynamic adaptation to complex and changing industrial conditions.

This paradigm can be implemented through two paths. The first is explicit modeling of indecisive fault-tolerant actions at knowledge boundaries. Through dedicated alignment training, lightweight agents learn to identify their own capability boundaries and choose fault-tolerant actions rather than giving forced answer or execution, when task parameters are missing, scenarios are unseen, or a task falls outside their knowledge scope [4,5]. The second is the DiSRouter (Distributed Self-Router) mechanism [14], which gives agents the ability to assess their own competence and autonomously decide whether to execute a task locally or delegate it upward to higher-capability agents, thereby avoiding execution failures caused by capability mismatch. Refusal and routing are representative fault-tolerant actions. Intent clarification, uncertainty annotation, and other diverse indecisive actions remain important research directions for single-agent fault-tolerance optimization.

## 5. Paradigm Two: Reliable Cross-Agent Communication via Semi-Formal Language Protocols

Single-agent local reliability does not guarantee reliable global collaboration across distributed agents. Current multi-agent collaboration frameworks embed all task constraints and handoff rules entirely within natural-language prompts, which produces vague and unverifiable governance rules. Minor semantic deviations at any single node can propagate and amplify along the collaboration chain and eventually lead to global task failure. This is a core challenge that cannot be addressed by single-node fault tolerance alone.

The root cause of agent communication errors is the inherent trade-off between natural language expressiveness and formal protocol verifiability. From the perspective of Chomsky's language hierarchy [19,20], agent communication always faces a trade-off between expressiveness and verifiability. Pure natural language is flexible but ambiguous; pure formal languages are rigorous but limited in semantic expressiveness for dynamic industrial scenarios. LLMs’ parsing performance also degrades with increasing prompt complexity. Reliable orchestration therefore requires external protocol-based control mechanisms, not just internal model capabilities.

We argue that reliable cross-agent collaboration should be built around an executable protocol programming system based on semi-formal language protocols [15]. Its central goal is to block error propagation along the communication chain and ensure system-level communication stability. The key design idea is to convert implicit rules embedded in natural-language prompts, such as task constraints, transition policies, and handoff responsibilities, into standardized formal protocol structures that are checkable, executable, and auditable. This separates explicit constraint auditing from implicit general reasoning while preserving the general semantic reasoning capability of natural language, thereby balancing expressive flexibility with machine verifiability.

This paradigm builds a dual fault-tolerance assurance mechanism for agent communication. At compile time, syntactic verification, logic checking, and parameter validation detect and block structural errors before execution. At runtime, guard mechanisms constrain agent interaction behavior in real time, block ambiguous information and invalid or rule-violating operations, and keep the collaborative process on reliable execution paths to prevent error propagation. This paradigm complements single-agent fault tolerance, addresses the systemic problem of "locally reliable but globally failing" collaboration, and supports stable long-horizon operation of DAS.

## 6. Industrial Deployment Challenges and Future Outlook

The distributed agent system framework is still in the transition from theory to industrial practice, with four core bottlenecks restricting large-scale deployment: missing standardized interfaces for heterogeneous agents; high overhead of existing fault-tolerant algorithms and protocol verification; absence of unified industrial standards for semi-formal communication protocols; and underdeveloped industrial reliability evaluation systems for long-horizon distributed agent coordination.

In response to these bottlenecks and in line with the development of trustworthy industrial AI, future research should focus on six directions: 1) dynamic reliability evolution mechanisms for heterogeneous end-device agents; 2) protocol-driven distributed collaboration deployment frameworks; 3) lightweight fault-tolerance and verification algorithms; 4) industry standards for semi-formal protocols; 5) open-source evaluation ecosystems for end-device agents; and multi-level fault-tolerance assurance for industrial safety [16,17,9].

Meanwhile, the theoretical framework proposed here is broadly adaptable. Beyond heterogeneous end-device scenarios, it can also be extended to long-horizon collaboration among multi-agent systems built around cloud-hosted frontier LLMs. Fault-tolerant collaboration under hybrid cloud-hosted deployments that combine small and large models is a particularly valuable future research direction [18].

## 7. Concluding Remarks

We introduce the Distributed Agent System (DAS) concept and propose a fault-tolerant collaboration paradigm for embodied agents based on the recognition that uncertainty is objectively unavoidable. It replaces the pursuit of single-turn zero error with system-level fault-tolerance management, thereby filling a critical gap between research on multi-agent systems built around cloud-hosted frontier LLMs and agent collaboration in constrained industrial end-device scenarios. By redefining agent reliability, it breaks with the traditional LLM-agent optimization logic which overemphasizes accuracy and error elimination. Single-agent fault-tolerant reliability alignment addresses errors at their source in individual end-device agents, while semi-formal language protocols address error propagation in multi-agent communication. Together, the two paradigms form an integrated fault-tolerance framework for trustworthy industrial AI. As lightweight algorithm design, protocol standardization, and evaluation-system standardization continue to advance, this framework can broadly support interconnection among intelligent end devices, industrial IoT, and other core scenarios, while providing new theoretical support and engineering pathways for reliable long-horizon multi-agent collaboration across broader application contexts.

## References

[1] Li X, Shi W, Zhang H, Peng C, Wu S, Tong W. The Agentic-AI Core: An AI-Empowered, Mission-Oriented Core Network for Next-Generation Mobile Telecommunications. Engineering 2026;56(1):104-119.

[2] Ma N, Pan J, Liu Y, Yang Y, Han Y, Guo J, Wu Z, Yang Z, Yang Z, Li D. Embodied Interactive Intelligence Towards Autonomous Driving. Engineering 2026;59(4):337-351.

[3] Fan D, Liu M, Shao Y, Yang L, Liu Y, Zhang Y, Ren Y, Wang Z. Domain-Specific Large Language Model for Maintenance Decision-Making on Wind Farms by Labeled-Data-Supervised Fine-Tuning. Engineering 2026;60(5):343-361.

[4] Xu H, Zhu Z, Pan L, Wang Z, Zhu S, Ma D, Cao R, Chen L, Yu K. Reducing tool hallucination via reliability alignment. In: Proceedings of the 42nd International Conference on Machine Learning; 2025 Jul 13-19; Vancouver, Canada. p. 69992-70006.

[5] Xu H, Zhu Z, Zhang S, Ma D, Fan S, Chen L, Yu K. Rejection improves reliability: training LLMs to refuse unknown questions using RL from knowledge feedback. In: Proceedings of the 1st Conference on Language Modeling; 2024 Oct 7-9; Philadelphia, PA, USA.

[6] Xu H, Yang Z, Zhu Z, Lan K, Wang Z, Wu M, Ji Z, Chen L, Fung P, Yu K. Delusions of large language models. 2025. arXiv:2503.06709.

[7] Liu Y, Zhou Y, Liu Y, Xu Z, He Y. Intelligent Fault Diagnosis for CNC Through the Integration of Large Language Models and Domain Knowledge Graphs. Engineering 2025;53(10):311-322.

[8] Chen J, Chen Y, Pham DT, Song Y, Wu J, Xing L, Chen Y. A Large Language Model-based Multi-Agent Framework to Autonomously Design Algorithms for Earth Observation Satellite Scheduling Problem. Engineering. In press.

[9] Camarinha-Matos LM. Collaborative Networks in Industry 4.0 and Industry 5.0. Engineering. In press.

[10] Chen Z, Yu W, Wu C, Chen F, Wang Z, Zhou C, You Y, Li S, Zhu Q, Ma N, Sun Y, Li D, Fanady B, Jiang S, Yan Z, Zhou S, Li L, Hsieh C, Bai Y, Xiao L, Chung C, Chan C, Cui Z, Grätzel M, Zhao H. Agentic Robotic Boxes for Perovskite Solar Cell Fabrication with Recipe Language Model. Engineering 2026;61(6):365-374.

[11] Piantadosi ST, Tily H, Gibson E. The communicative function of ambiguity in language. Cognition 2012;122(3):280-291.

[12] Zipf GK. Human Behavior and the Principle of Least Effort: An Introduction to Human Ecology. Addison-Wesley Press, 1949.

[13] Ren J, Sun Y, Du H, Yuan W, Wang C, Wang X, Zhou Y, Zhu Z, Wang F, Cui S. Generative Semantic Communication: Architectures, Technologies, and Applications. Engineering 2026;56(1):45-61.

[14] Zheng H, Xu H, Lin YK, Fan S, Chen L, Yu K. DiSRouter: distributed self-routing for LLM selections. In: Proceedings of the 14th International Conference on Learning Representations; 2026 Apr 23-27; Rio de Janeiro, Brazil.

[15] Li H, Peng J, Wang Z, Chen L, Yu K. XFlow: an executable protocol programming system for reliable multi-agent workflows. 2026. arXiv:2606.14790.

[16] Huang K, Liang C, Hua M, Zhan X, Ge M. Hierarchical three-dimensional output tracking for networked uncertain robotic systems with Byzantine fault-tolerant capability. Nonlinear Dynamics 2026, 114(9):627.

[17] Wu Q, Du S. Fixed-time fault-tolerant consensus of uncertain multiagent systems via fully distributed control. IET Control Theory & Applications 2026;20(1):e70123.

[18] Wu F, Shen T, Bäck T, Chen J, Huang G, Jin Y, Kuang K, Li M, Lu C, Miao J, Wang Y, Wei Y, Wu F, Yan J, Yang H, Yang Y, Zhang S, Zhao Z, Zhuang Y, Pan Y. Knowledge-Empowered, Collaborative, and Co-Evolving AI Models: The Post-LLM Roadmap. Engineering 2025;44:87-100.

[19] Chomsky N. Three models for the description of language. IRE Transactions on Information Theory 1956;2(3):113-124.

[20] Chomsky N. The Minimalist Program. Cambridge, MA: MIT Press; 2014.

[21] Young S. Cognitive User Interfaces. IEEE Signal Processing Magazine 2010;27(3):128-140.

[22] Young S, Gasić M, Keizer S, Mairesse F, Schatzmann J, Thomson B and Yu K. The Hidden Information State Model: a practical framework for POMDP-based spoken dialogue management. Computer Speech and Language 2010;24(2):150-174.